\documentclass[%
reprint,
superscriptaddress,
amsmath,amssymb,
aps,
prl
]{revtex4-2}

\usepackage{graphicx}
\usepackage{dcolumn}
\usepackage{bm}
\usepackage[colorlinks=true, letterpaper=true, pdfstartview=FitV, linkcolor=blue, citecolor=blue, urlcolor=blue]{hyperref}

\begin{document}

\preprint{APS/123-QED}

\title{Giant spin shift current in two-dimensional altermagnetic multiferroics VOX$\mathrm{_2}$ }

\author{Yao Yang}
\email{yangyao@szu.edu.cn}
\affiliation{ College of Physics and Optoelectronic Engineering, Shenzhen University, Shenzhen 518060, China }

\date{\today}

\begin{abstract}

Altermagnets represent a novel class of magnetic materials that integrate the advantages of both ferromagnets and antiferromagnets, providing a rich platform for exploring the physical properties of multiferroic materials. 
 This work demonstrates that $\mathrm{VOX_2}$  monolayers ($\mathrm{X = Cl, Br, I}$) are two-dimensional ferroelectric altermagnets, as confirmed by symmetry analysis and first-principles calculations. 
$\mathrm{VOI_2}$ monolayer exhibits a strong magnetoelectric coupling coefficient ($\alpha_S \approx 1.208 \times 10^{-6}~\mathrm{s/m}$), with spin splitting in the electronic band structure tunable by both electric and magnetic fields. 
Additionally, the absence of inversion symmetry in noncentrosymmetric crystals enables significant nonlinear optical effects, such as shift current (SC). 
 The $x$-direction component of SC exhibits a ferroicity-driven switching behavior.
Moreover, the $\sigma^{yyy}$ component  exhibits an exceptionally large spin SC of $330.072~\mathrm{\mu A/V^2}$. 
These findings highlight the intricate interplay between magnetism and ferroelectricity, offering versatile tunability of electronic and optical properties. 
$\mathrm{VOX_2}$ monolayers provide a promising platform for advancing two-dimensional multiferroics, paving the way for  energy-efficient memory devices, nonlinear optical applications and opto-spintronics.

\end{abstract}

\maketitle

\section{Introduction}

Magnetoelectric multiferroics are materials that simultaneously exhibit ferromagnetic and ferroelectric properties, with an intrinsic coupling between these two order parameters \cite{Cross_1987_Relaxor}. 
The ability to control magnetism using an electric field, or vice versa, holds great potential for applications in low-power electronics, sensors, memory devices, and emerging computing technologies \cite{Eerenstein_2006_Multiferroic,Fiebig_2016_Evolution,Lu_2019_Singlephase,Spaldin_2019_Advances,Yang_2022_Twodimensional}. 
The discovery and development of magnetoelectric multiferroics have unlocked exciting opportunities in materials science and condensed matter physics.

On one hand, due to the intrinsic repulsion between charge polarization and spin polarization, multiferroic materials are inherently rare \cite{Khomskii_2006_Multiferroics, Wang_2009_Multiferroicity, Lee_2010_Strong, Dong_2019_Magnetoelectricity}. 
On the other hand, conventional magnetic materials are typically classified as either ferromagnetic or antiferromagnetic. 
Compared to ferromagnets, antiferromagnets offer higher information storage density and unique terahertz spin dynamics, enabling magnetic moment reversal on a picosecond timescale \cite{Baltz_2018_Antiferromagnetic, Smejkal_2018_Topological, Shao_2024_Antiferromagnetic, Chen_2024_Emerging}. 
However, the weak response of antiferromagnets to external fields and the challenges in controlling their spin order significantly limit their application in multiferroic systems.

Recently, altermagnets have been proposed as a third type of non-relativistic collinear magnetic phase, distinct from both ferromagnets and antiferromagnets \cite{Smejkal_2022_Conventional,Smejkal_2022_Emerging}. 
Altermagnets combine the advantages of both ferromagnets and antiferromagnets, exhibiting zero net magnetization while maintaining momentum-dependent spin splitting \cite{Bai_2024_Altermagnetism}. 
It has been experimentally confirmed in materials such as $\mathrm{MnTe}$ \cite{Krempaský_2024_Altermagnetic,Lee_2024_Broken} and $\mathrm{CrSb}$ \cite{Yang_2025_Threedimensional,Ding_2024_Large}. 
The discovery of altermagnets not only circumvents the limitations of antiferromagnetic multiferroics but also introduces a new paradigm for magnetoelectric multiferroic materials.
The emergence of altermagnets has brought about many novel physical phenomena, such as unique spin currents \cite{González-Hernández_2021_Efficient}, giant magnetoresistance, tunnel magnetoresistance \cite{Smejkal_2022_Giant,Cui_2023_Giant,Samanta_2024_Tunneling}, 
the anomalous Hall effect \cite{GonzalezBetancourt_2023_Spontaneous}, and  the quantum anomalous Hall effect \cite{Guo_2023_Quantum}.

With the rapid development of two-dimensional (2D) materials, exemplified by graphene, an increasing number of novel layered structures have been explored \cite{Novoselov_2004_Electric,Mak_2010_Atomically,Li_2014_Black}. 
The advent of 2D materials provides a promising pathway for the miniaturization and integration of electronic components. 
Thus, achieving the coexistence of ferroelectricity and magnetism in 2D materials has become a significant research direction in the field of magnetoelectric multiferroics. 
Yang et al. first proposed the concept of sliding ferroelectricity as a means to achieve electric field control of altermagnetism \cite{Sun_2024_Altermagnetism}. 
Zhou et al. further demonstrated the control of altermagnetism via an electric field in antiferroelectric altermagnets, providing experimental validation of magnetoelectric coupling \cite{Duan_2025_Antiferroelectric}. 
Despite the advantages of altermagnets, the integration of ferroelectricity into multiferroic systems to control altermagnetism remains underexplored. 
Further investigations in this field are crucial for advancing next-generation multiferroic materials.

 This work proposes $\mathrm{VOX_2}$  ($\mathrm{X = Cl, Br, I}$) as two-dimensional ferroelectric altermagnets based on symmetry analysis. 
First-principles calculations reveal that $\mathrm{VOX_2}$ exhibits an electronic structure with spin splitting and huge magnetoelectric coupling. 
Further studies indicate that the band splitting can be effectively tuned by electric and magnetic fields. 
Additionally, the $\sigma^{xxx}$ component of $\mathrm{VOX_2}$ is tunable via ferroelectric polarization. 
Moreover,  the $\sigma^{yyy}$ component of the $\mathrm{VOI_2}$ monolayer exhibits an exceptionally large spin shift current (SC).
This study not only deepens the understanding of magnetoelectric coupling mechanisms in two-dimensional materials but also holds profound implications for the advancement of  energy-efficient memory devices, nonlinear optical applications, and opto-spintronic technologies.

\begin{figure*}[t]
  \includegraphics[width=0.85\linewidth]{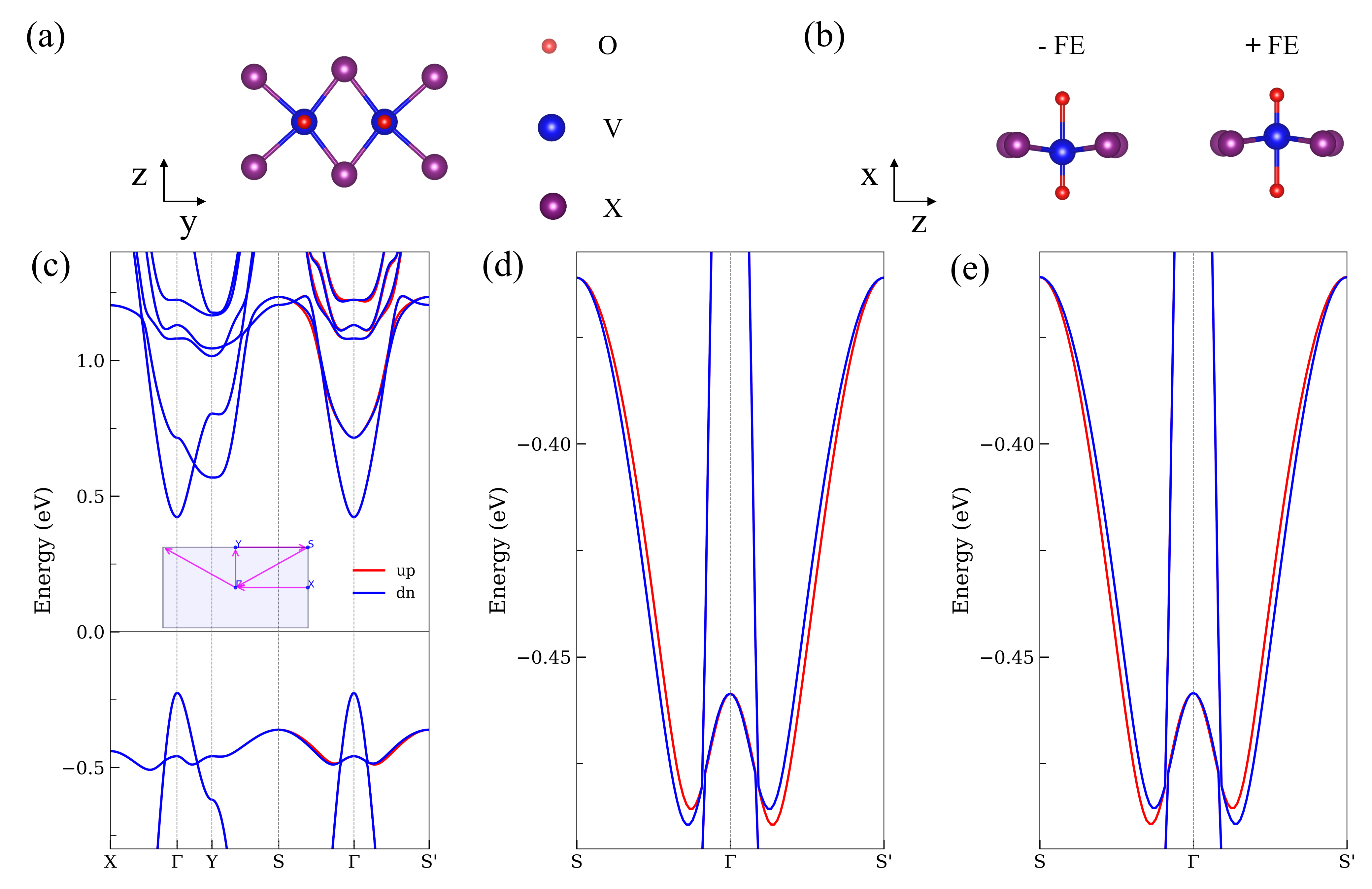}
  \caption{\label{fig:Crystal_Band}   (a)  The side view of altermagnetic  phase $\mathrm{VOI_2}$ monolayer, 
  (b)  The top view of negative  ($P_\downarrow$) and positive  ($P_\uparrow$) of altermagnetic phase $\mathrm{VOI_2}$  monolayer,
  (c) The spin-polarized band structure of $\mathrm{VOI_2}$,
  (d) and (e) The spin-projected valence band of  $P_\downarrow$  and  $P_\uparrow$  ferroelectric configurations.
  }
  \end{figure*}

\section{Crystal Structure and Symmetry Analysis}

$\mathrm{VOI_2}$ belongs to the van der Waals (vdW) multiferroic family $\mathrm{VOX_2}$  \cite{Xu_2024_Firstprinciples}. 
The  $\mathrm{VOI_2}$ monolayer forms a two-dimensional octahedral network, where adjacent V atoms share a corner oxygen along the $a$ direction and an iodine  edge along the $b$ direction. 
First-principles calculations predict that the $\mathrm{VOI_2}$ monolayer exhibits both ferroelectricity and ferromagnetism \cite{Tan_2019_Twodimensional}, with the V atoms at the center of the octahedra displaced toward the oxygen atoms. 
This configuration is referred to as the FE-\uppercase\expandafter{\romannumeral1} phase. 
Peierls transitions frequently occur in quasi-one-dimensional chain-like structures \cite{Gooth_2019_Axionic, Zhang_2020_Firstprinciples}. 
The $\mathrm{VOI_2}$ monolayer undergoes a V-V dimerization, leading to a Peierls transition, forming the FE-\uppercase\expandafter{\romannumeral2} phase \cite{Zhang_2021_Peierls}. 
This phase belongs to the space group $\mathrm{Pmm2}$ (No. 25), as shown in Fig. \ref{fig:Crystal_Band}(a). 
The two ferroelectric polarization directions  ($P_{\downarrow}$ and $P_{\uparrow}$) are illustrated in Fig. \ref{fig:Crystal_Band}(b), respectively.

Based on crystal field theory, the V 3$d$ orbitals in $\mathrm{VOX_2}$ monolayer under an octahedral environment.
$\mathrm{VOX_2}$ has a $d^1$ configuration, where $\mathrm{V^{4+}}$ hosts a single electron in the $d_{xy}$ orbital. 
The spin charge density of $\mathrm{VOX_2}$ monolayer, shown in Fig. S2, reveals a strong $d_{xy}$ orbital character. 
The interactions between V-$3d_{xy}$ orbitals along the $b$ axis  lead to V-V dimers, forming alternating $\sigma$ bonding and $\sigma^{\ast}$ antibonding states. 
The exchange interaction along the V-X direction is governed by the competition between two types of exchange mechanisms \cite{Tan_2019_Twodimensional}: 
(1) the direct exchange interaction between neighboring V atoms, favoring antiparallel spin alignment, and 
(2) the superexchange interaction mediated by halogen atoms, promoting parallel spin alignment. 
The Peierls transition enhances the direct exchange interaction, favoring antiparallel spin arrangements, consistent with the Goodenough-Kanamori-Anderson rule \cite{jin_2024_anomalous}. 
$\mathrm{VOX_2}$ monolayer exhibits parallel spin alignment along the V-O direction, akin to the FE-\uppercase\expandafter{\romannumeral1} phase \cite{Tan_2019_Twodimensional}.

\section{Electronic Structure}


 The ferroelectric polarization in $\mathrm{VOI_2}$ breaks the $\mathcal{P}$ symmetry, 
and the Peierls transition eliminates the $\tau_{\frac{1}{2}y}$ symmetry. 
Compared with the undistored structure, 
the altermagnetic order breaks both $\mathcal{P}\mathcal{R}_{s}$ and $\mathcal{R}_{s}\tau_{\frac{1}{2}y}$ symmetries, where $\mathcal{R}_{s} \tau_{\frac{1}{2}y}$ denotes the combined operation of spin-reversal and half-lattice translation along $y$ axis. 
This unconventional symmetry breaking permits intrinsic spin splitting at arbitrary crystal momentum $\mathbf{k}$ even in the absence of SOC.
Fig. \ref{fig:Crystal_Band}(c) presents the band structure of monolayer $\mathrm{VOI_2}$ along both high-symmetry and general \textbf{k} paths.
Notably, along high-symmetry \textbf{k} paths, the bands exhibit Kramers spin degeneracy, particularly along the X-$\Gamma$-Y-S path. 
While the magnitude of spin splitting adheres to the full nonmagnetic crystal symmetry, its sign alternates, ensuring that the material as a whole maintains spin-compensated symmetry \cite{jin_2024_anomalous}.
As a result, the net magnetization of the altermagnetic phase of monolayer $\mathrm{VOI_2}$ remains zero, as clearly indicated in the density of states in Fig. S2(c). 
 The results reveal that the first valence and conduction bands exhibit spin splittings with maximum values of 12.26 meV and 37.68 meV, respectively.
The spin splitting energy is comparable to that of $\mathrm{SnS_2/MnPSe_3/SnS_2}$ heterostructures (19.1 meV).

The $\mathrm{VOI_2}$ monolayer exhibits both altermagnetism and ferroelectricity. 
It possesses four distinct configurations, denoted as  $P_{\uparrow} M_{\uparrow\downarrow}$, $P_{\uparrow} M_{\downarrow\uparrow}$, $P_{\downarrow} M_{\uparrow\downarrow}$, and $P_{\downarrow} M_{\downarrow\uparrow}$.
Here, $P_\uparrow$ indicates ferroelectric polarization along the positive $a$-axis direction, while $M_{\uparrow\downarrow}$ denotes the magnetic moments of the left (right) V atom oriented along the positive (negative) $a$-axis direction.
The calculations reveal that all four ferroelectric altermagnetic configurations are energetically degenerate. 
However, the alternating spin splitting varies with the direction of ferroelectric polarization and the magnetic configuration of V atoms.

The configurations  $P_{\uparrow} M_{\downarrow\uparrow}$ and $P_{\downarrow} M_{\uparrow\downarrow}$ exhibit identical spin-alternating band structures, as shown in Fig. \ref{fig:Crystal_Band}(d). 
Similarly,  $P_{\uparrow} M_{\uparrow\downarrow}$ and $P_{\downarrow} M_{\downarrow\uparrow}$  display the same spin-split bands, as illustrated in Fig. \ref{fig:Crystal_Band}(e). 
The magnitude of spin splitting in the band structure remains equal but exhibits opposite signs for the  $P_{\uparrow} M_{\uparrow\downarrow}$ and $P_{\uparrow} M_{\downarrow\uparrow}$ configurations.
 The band structures with spin orbit coupling are shown in Fig. S5 (a) and (b).
This phenomenon is similar to  multiferroic bilayer $\mathrm{VS_2}$\cite{Liu_2020_Magnetoelectric}.

The ability to control the altermagnetic order in $\mathrm{VOX_2}$ via ferroelectric polarization highlights its significant potential for next-generation high-performance electronic devices. 
Furthermore, the altermagnetic structure of V atoms can be switched by applying an external magnetic field \cite{Liu_2020_Magnetoelectric}. 
By leveraging both external magnetic and electric fields, arbitrary switching among all four distinct ferroelectric interlaced magnetic states can be achieved, opening new possibilities for multistate memory applications.

\section{Magnetoelectric Coupling}

\begin{figure}[t]
  \centering
  \includegraphics[width=1\linewidth]{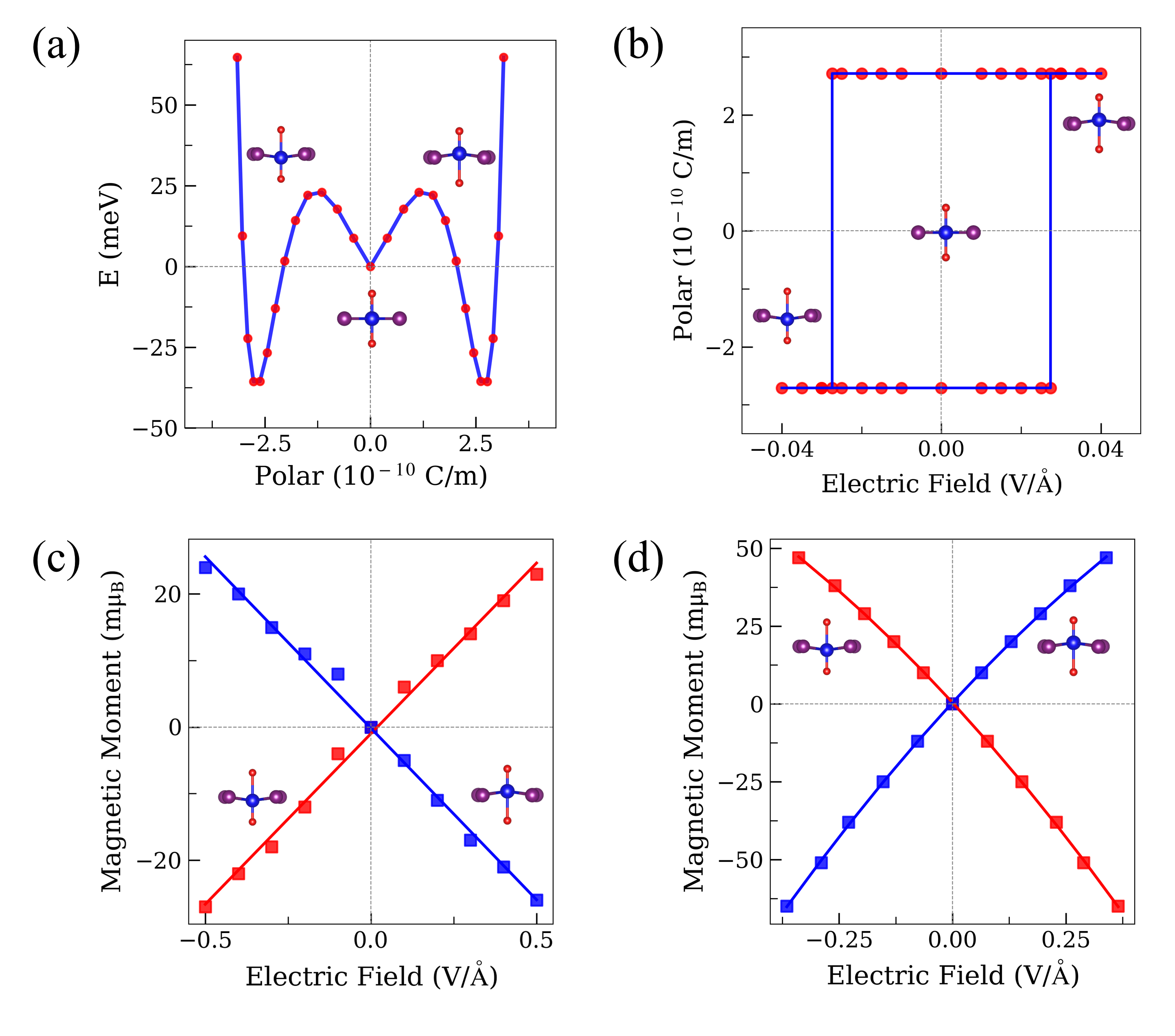}
  \caption{\label{fig:MagnetoelectricCoupling} 
  (a) The relationship between polarization intensity and energy during the ferroelectric phase transition. 
  (b) The polarization-electric field (P-E) hysteresis loop illustrating ferroelectric switching under an applied external electric field. 
  (c) The electronic magnetoelectric coupling relationship. 
  (d) The ionic magnetoelectric coupling relationship.}
\end{figure}

Controlling magnetism via an electric field is one of the key challenges in next-generation information technology. 
To investigate the ferroelectric properties of the altermagnetic phase, 
 the polarization of VOX2 monolayers is calculated using the Berry phase method \cite{King-Smith_1993_Theory,Resta_1993_Quantum}. 
The ferroelectric polarization of $\mathrm{VOI_2}$ in the  $P_{\downarrow} M_{\uparrow\downarrow}$ configuration along the in-plane $a$-axis direction is $-2.77\times 10^{-10}~\mathrm{C/m}$, which is close to previous calculations \cite{Tan_2019_Twodimensional}. 
The corresponding polarization field is estimated as $E_P \approx 7.026~\mathrm{V/}$\AA ~using the relation $E = P/\epsilon_0$. 

To gain further insight into the ferroelectric transition, 
 the switching pathway is simulated by linearly interpolating atomic positions between different ferroelectric structures. 
The energy barrier for the phase transition in the altermagnetic monolayer $\mathrm{VOI_2}$ is calculated as 35.643 meV, as shown in Fig. \ref{fig:MagnetoelectricCoupling}(a). 
The transition barrier increases as the number of halogen elements decreases, as shown in Fig. S7.
The polarization-electric field (P-E) hysteresis loop characterizes the response of ferroelectric materials to an applied electric field \cite{Neugebauer_1992_Adsorbatesubstrate}, representing a fundamental feature of ferroelectricity.
The calculations indicate a coercive field of 27.3 mV/\AA, as depicted in Fig. \ref{fig:MagnetoelectricCoupling}(b), demonstrating that ferroelectric polarization switching can be achieved with a relatively small external electric field.

The $\mathrm{VOX_2}$ monolayer is predicted to exhibit strong magnetoelectric coupling \cite{Tan_2019_Twodimensional}. 
The magnetoelectric coupling coefficient can be classified into two categories: electronic and ionic magnetoelectric coupling \cite{Xu_2024_Firstprinciples}. 
The variation of net magnetization with an applied electric field exhibits an approximately linear dependence, as shown in Fig. \ref{fig:MagnetoelectricCoupling}(c). 
This relationship follows the expression $\mu_0 \Delta M = \alpha_S E$ \cite{Xu_2024_Firstprinciples}, where $\alpha_S$ is the linear magnetoelectric coupling coefficient. 
Here, the positive electric field is defined along the positive $a$-axis direction. 
By performing a linear fit to the data, $\alpha_S \approx 3.915 \times 10^{-7}~\mathrm{s/m}$ for the  $P_{\downarrow} M_{\uparrow\downarrow}$ configuration and $\alpha_S \approx -3.936 \times 10^{-7}~\mathrm{s/m}$ for the  $P_{\uparrow} M_{\uparrow\downarrow}$ configuration.

Furthermore, during the ferroelectric phase transition, the magnetic moment undergoes a corresponding change due to the displacement of ions. 
The ionic magnetoelectric coupling originates from the relative positions of V ions and is closely correlated with ferroelectric polarization. 
Fig. \ref{fig:MagnetoelectricCoupling}(d) illustrates that the magnetic moment exhibits a nonlinear dependence on the applied electric field. 
This behavior can be numerically described by the expression $\mu_0 \Delta M = \beta_S E^2 + \alpha_S E$,
where $\beta_S$ represents the second-order nonlinear magnetoelectric coefficient \cite{Xu_2024_Firstprinciples}. 
For the  $P_{\downarrow} M_{\uparrow\downarrow}$ configuration,  $\beta_S \approx -4.494 \times 10^{-17}~\mathrm{s/m}$ and $\alpha_S \approx -1.208 \times 10^{-6}~\mathrm{s/m}$. 
In the  $P_{\uparrow} M_{\uparrow\downarrow}$ configuration, the coefficients are $\beta_S \approx -4.494 \times 10^{-17}~\mathrm{s/m}$ and $\alpha_S \approx 1.208 \times 10^{-6}~\mathrm{s/m}$. 

\begin{figure*}[htbp]
  \centering
  \includegraphics[width=0.95\linewidth]{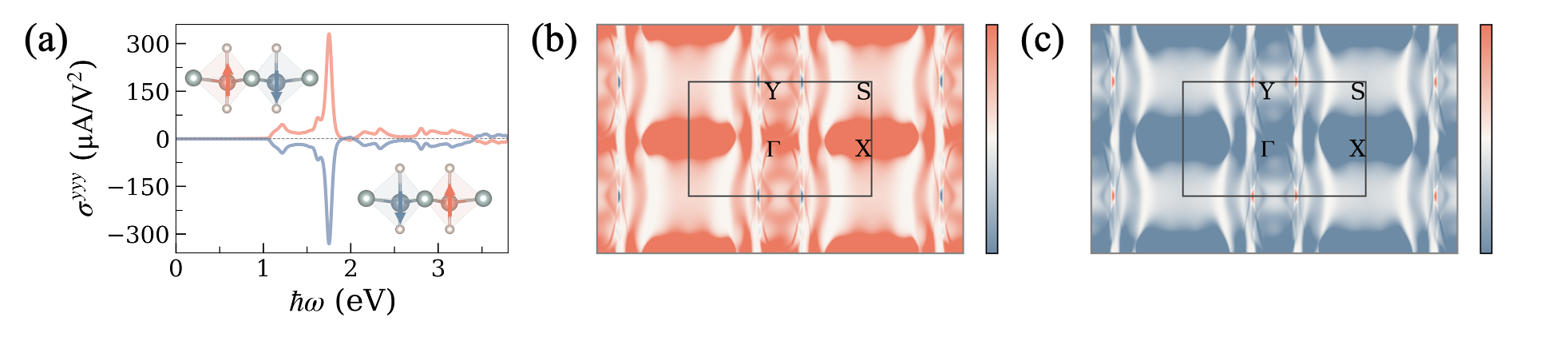}
  \caption{\label{fig:SpinSC}
  (a) Spin shift current spectra for the $P_{\downarrow} M_{\uparrow\downarrow}$ and $P_{\downarrow} M_{\downarrow\uparrow}$ configurations, demonstrating reversal of the photocurrent upon magnetic order switching.  
  (b), (c) Momentum-resolved distributions of the spin-up and spin-down components of the spin shift current under the $P_{\downarrow} M_{\uparrow\downarrow}$ configuration, respectively.  
  The antisymmetric distribution of opposite-spin contributions confirms the generation of pure spin photocurrents in the absence of net charge current.
  }
\end{figure*}

The $\mathrm{VOX_2}$ monolayer exhibits both a substantial electronic magnetoelectric coupling effect and an unparalleled ionic magnetoelectric coupling effect. 
These coupling strengths are significantly larger than those observed in iron thin films ($\alpha_S^{001} = 4.35 \times 10^{-8}~\mathrm{s/m}$) \cite{Duan_2008_Surface}. 
The exceptionally large magnetoelectric coupling coefficient suggests a promising avenue for electric-field control of magnetism, with potential applications in spintronics, low-power memory devices, tunable optics, and quantum computing.

\section{ Shift Current}

The absence of inversion symmetry in the $\mathrm{VOX_2}$ monolayer gives rise to a SC, 
which is a nonlinear DC photocurrent induced by light \cite{Wang_2017_Firstprinciples,Ibañez-Azpiroz_2018_Initio,Zhang_2019_Enhanced}. 

 Fig. S7 illustrate the relationship between the SC susceptibility tensor $\sigma^{xxx}$ and photon energy for the $\mathrm{VOI_2}$ monolayer. 
The $\sigma^{xxx}$ component exhibits four distinct peaks in the 0–3.5 eV range.
The first peak of $\sigma^{xxx}$ appears at $h\omega_0 = 1.180~\mathrm{eV}$, with a value of $4.443~\mathrm{\mu A/V^2}$. 
This contribution primarily originates from \textbf{k}-points near the center of the Brillouin zone (the $\Gamma$-point), as shown in  Fig. S8. 
The highest peak at $h\omega_0 = 3.150~\mathrm{eV}$ exhibits a comparable magnitude of $39.775~\mathrm{\mu A/V^2}$, arising from transitions at \textbf{k}-points near the Brillouin zone boundary (X point) (Fig. S8).
This SC magnitude is comparable to that of typical two-dimensional materials, such as $\mathrm{2H-MoS_2}$ ($8~\mathrm{\mu A/V^2}$) \cite{Schankler_2021_Large}. 

Since SC is a polar vector, its sign reverses upon ferroelectric polarization switching ($P_x\to -P_x$) \cite{Wang_2019_Ferroicitydriven}. 
Consequently, the sign of the SC susceptibility tensor $\sigma^{xxx}$ also inverts.
Thus, under the same linearly polarized light, the SC direction undergoes a 180° shift upon ferroelectric polarization switching, following the relation:
$J_{\mathrm{SC}}^{x, \leftrightarrow}\left(P_x\right)=-J_{\mathrm{SC}}^{x, \leftrightarrow}\left(-P_x\right)$, $J_{\mathrm{SC}}^{x, \updownarrow}\left(P_x\right)=-J_{\mathrm{SC}}^{x, \updownarrow }\left(-P_x\right)$.
 Fig.S7 presents the SC susceptibility tensor $\sigma^{xxx}$ after ferroelectric switching in the $\mathrm{VOI_2}$ monolayer. 
The peak of $\sigma^{xxx}$ remains at $h\omega_0 = 1.890~\mathrm{eV}$, but its magnitude is equal in absolute value and opposite in sign compared to its pre-transition value ($28.00~\mathrm{\mu A/V^2}$). 
This result demonstrates that external electric fields can effectively regulate SC via ferroelectric switching \cite{Wang_2020_Electrically,Zhang_2022_Tailoring}. 
Ferroelectric control of SC provides an excellent platform for the development of nonlinear optoelectronics in multiferroic semiconductors.

Moreover, 
For the  $P_{\downarrow} M_{\uparrow\downarrow}$ configuration, the $\sigma^{yyy}$ component exhibits two prominent peaks at photon energies of 1.755 eV and 1.255 eV, as shown in  Fig. \ref{fig:SpinSC}(a). 
The largest peak occurs at 1.755 eV, reaching a substantial magnitude of $330.07~\mathrm{\mu A/V^2}$, while the secondary peak at 1.255 eV has a value of $44.66~\mathrm{\mu A/V^2}$. 
To elucidate the origin of these peaks, the \textbf{k}-resolved contributions of the spin SC  are analyzed. 
 The dominant peak at 1.755 eV arises from transitions near the $\Gamma$-point and the X-point (Figs. \ref{fig:SpinSC}(b) and (c))
whereas the first peak at 1.255 eV primarily originates from electronic transitions near the $\Gamma$-point(Fig. S9).
Furthermore, for the reversed  ferroelectric altermagnetic polarization  ($P_{\uparrow} M_{\downarrow\uparrow}$), the peak value of the spin SC at 1.755 eV is equal in magnitude but opposite in sign ($-330.07~\mathrm{\mu A/V^2}$)  (Fig. \ref{fig:SpinSC}(a)). 
This behavior clearly demonstrates the ferroelectric tunability of the spin SC. 
The remarkable magnitude and switchability of both charge and spin SC highlight the potential of $\mathrm{VOI_2}$ monolayers as promising candidates for nonlinear optoelectronic and optospintronic devices.

\subsection{Discussion}

This work investigates the unique properties of the altermagnetic phase in $\mathrm{VOX_2}$  monolayers, highlighting their ferroelectric and altermagnetic multiferroic characteristics. 
Large quantities of oxide dichlorides $\mathrm{MOX_2}$ have been studied for structural phase transitions\cite{Jia_2019_Niobium,Zhang_2021_Peierls,Zhang_2021_Orbitalselective,Guo_2023_Ultrathin}, which can break the  $\mathcal{P}$ and $\mathcal{\tau}_{\frac{1}{2}}$ symmetries.
Altermagnetic phase may also exist in the vdW family of layered oxide dichlorides $\mathrm{MOX_2}$.

The absence of inversion symmetry in these materials enables the emergence of nonlinear optical effects, such as SC.
 In addtion, ferroelectric switching induces a 180° reversal of the SC. 
Other nonlinear optical effects and polar vector-related phenomena, such as injection currents, can also be driven by ferroelectricity \cite{Wang_2019_Ferroicitydriven}.

\section{ Conclusion}

In summary, the ferroelectric altermagnetism in $\mathrm{VOX_2}$  monolayer  is investigated through symmetry analysis and first-principles calculations.
 The results demonstrate that $\mathrm{VOX_2}$  monolayer represent a novel class of two-dimensional ferroelectric altermagnets.
Further studies reveal that external fields (electric and magnetic) can reverse the band splitting, highlighting a significant magnetoelectric coupling effect.
Furthermore, the nonlinear optical response, as a fundamental property of non-centrosymmetric materials, is explored.
 The findings indicate that $\mathrm{VOI_2}$ exhibits a pronounced SC and the SC along the $x$-direction can be reversed via ferroelectric control.
Ferroelectric modulation of the SC presents a practical strategy for controlling nonlinear optical effects, establishing an excellent platform for exploring multiferroic semiconductors and advancing the development of high-performance optoelectronic devices.
Moreover, an exceptionally large spin SC under $y$-direction linearly polarized light.
This proves the feasibility of controlling magnetic states by electromagnetic waves in non-centrosymmetric materials and provides a new opportunity for the development of opto-spintronic applications.
This work opens new avenues for investigating the interplay between ferroelectricity and altermagnetism in two-dimensional materials, laying the foundation for future innovations in energy-efficient electronics  and opto-spintronics.

Note added: During the revision of this manuscript, the author noticed  a recent work by Z. Zhu et al.\cite{Zhu_2025_Twodimensional}, which proposed a wide range of ferroelectric altermagnets, including $\mathrm{VOI_2}$ studied in the present work.

\section{Acknowledgments}
\begin{acknowledgments}
    Y. Yang acknowledges support from the Physics Postdoctoral Science Funding of Shenzhen University.
\end{acknowledgments}

\nocite{*}

\bibliography{apssamp}

\end{document}